\documentclass[superscriptaddress,showpacs,aps,prl,twocolumn]{revtex4}

\usepackage{graphicx}
\usepackage{amssymb}

\def\smsr{Sm$_{1-x}$Sr$_x$MnO$_3$}

\def\smsrA{Sm$_{0.55}$Sr$_{0.45}$Mn$^{18}$O$_3$}

\def\smsrC{Sm$_{0.5}$Sr$_{0.5}$Mn$^{18}$O$_3$}

\def\smsrAC{Sm$_{1-x}$Sr$_{x}$Mn$^{18}$O$_3$}

\def\ndeusr{(NdEu)$_{0.55}$Sr$_{0.45}$Mn$^{18}$O$_3$}
\def\osix{$^{16}$O}
\def\oeight{$^{18}$O}
\def\fref#1{Fig.~\ref{#1}}

\begin{document}

\title{Quenched disorder induced magnetization jumps in (Sm,Sr)MnO$_3$}

\author{L. M. Fisher}
\affiliation{All-Russian Electrical Engineering Institute, 12 Krasnokazarmennaya Street, 111250 Moscow, Russia}
\author{A. V. Kalinov}\email[Corresponding author. E-mail: ]{kalinov@vei.ru}
\affiliation{All-Russian Electrical Engineering Institute, 12 Krasnokazarmennaya Street, 111250
Moscow, Russia} \affiliation{Materials Science Center, University of Groningen, Nijenborgh 4, 9747
AG Groningen, The Netherlands}
\author{I. F. Voloshin}
\affiliation{All-Russian Electrical Engineering Institute, 12 Krasnokazarmennaya Street, 111250 Moscow, Russia}
\author{N. A. Babushkina}
\affiliation{Institute of Molecular Physics, Russian Research Center ``Kurchatov Institute'',
Kurchatov Square 1, 123182 Moscow, Russia}
\author{D. I. Khomskii}
\affiliation{Materials Science Center, University of Groningen, Nijenborgh 4, 9747 AG Groningen,
The Netherlands} \affiliation{II Physikalisches Institut, Universit\"at zu K\"oln, Z\"ulpicher
Str.\ 77, 50937 K\"oln, Germany}
\author{Y. Zhang}
\affiliation{Department of Physics, Leiden University, Niels Bohrweg 2, 2300 RA Leiden, The
Netherlands}
\author{T. T. M. Palstra}
\affiliation{Materials Science Center, University of Groningen, Nijenborgh 4, 9747 AG Groningen,
The Netherlands}

\begin{abstract}
Magnetic field induced step-like changes in magnetization and resistivity of \smsr\ manganites were
studied. A strong dependence of these features on the cooling rate was observed. Magnetostriction,
however, does not show the presence of large strain in our samples. From all these features we can
rule out the conventional explanation of magnetization jumps as a consequence of martensitic
transition. We propose instead that quenched by fast cooling disorder leads to the formation of an
inhomogeneous metastable state and to subsequent magnetization jumps.
\end{abstract}

\date{\today}

\maketitle

%($\vartriangle$) ($\triangledown$) ($\square$) ($\blacksquare$) ({\Large $\circ$})

% >>>>>>>>>>>  INTRO  <<<<<<<<<<<<<<<<<<

Magnetic field-induced first order phase transitions attract a lot of attention both in
conventional antiferromagnets (AFM) \cite{metamagnetism} and in mixed-valence manganites (see
Ref.~\onlinecite{CoeyReview} and references therein) as well as in some pseudobinary systems
\cite{HardyGdGe, LevinGdGe}. In AFM these transitions are usually reversible and relatively broad
\cite{metamagnetism}. In diluted metamagnets (for example, Fe$_x$Mg$_{1-x}$Cl$_2$) they may be
steep (avalanche-like) and hysteretic \cite{KushauerFeMgCl}. In manganites such transitions may be
sharp or broad, reversible or strongly hysteretic \cite{XiaoJAP97} depending on chemical
composition and temperature, and are often accompanied by structural and insulator-to-metal (I-M)
transition \cite{TomiokaCOcollapse, TomiokaMFIndMIT, YoshizawaNeutrFI_MIT}.

Recently, the field-induced phase transition to a ferromagnetic (FM) state was shown to be
discontinuous at low $T < 5$~K in ceramic Mn-doped Pr$_{0.5}$Ca$_{0.5}$MnO$_3$ \cite{HebertStepsNi,
HebertAvalDoped}, in ceramics and single crystal of Pr$_{1-x}$Ca$_x$MnO$_3$ ($x = 0.3 - 0.37$)
\cite{MahendiranStepsPRL, HardyJMMM} and in Gd$_5$(SiGe)$_4$ alloys \cite{LevinGdGe, HardyGdGe}.
The position and number of steps depend on the magneto-temperature history and on the magnetic
field sweep rate \cite{HardyGdGe}. This was interpreted as the result of some kind of martensitic
transformation. However, this scenario is not clear because grain boundaries in ceramics could be
intrinsic barriers for domain-wall movement. Ghivelder et al.\ \cite{LevyMagnetocalor} have
observed a huge temperature increase at such abrupt field-induced transition (from 2.5 to 30~K),
whereas the specific heat before and after transition differs only by 10\%. This implies that a
large (magnetic) entropy is frozen in the sample and abruptly released upon increase of the
magnetic field. The authors of Ref.~\onlinecite{LevyMagnetocalor} have proposed a model in which
the local AFM-FM transition releases the heat locally and triggers a heat avalanche leading to the
observed magnetization jumps. In this case, the step should have some finite characteristic
time-scale of the order of a thermal relaxation time. However, only the magnetic field width of the
step was discussed and was found to be less than 2~Oe \cite{MahendiranStepsPRL} or even strictly
zero \cite{LevyMagnetocalor}.

In this Letter we show that step-like behavior exists in the magnetization $M$ and resistivity
$\rho$ but not in the magnetostriction of \smsrAC\ ($x = 0.45; 0.5$) ceramics. These steps have a
characteristic time-scale of the order of 1~ms which does not depend on the magnetic field sweep
rate. Moreover, the low-field low-temperature magnetic state itself strongly depends on the
zero-field cooling rate. There are no $M$ and $\rho$ steps for slowly (1~K/min) cooled samples, but
they exist only for rapidly cooled samples. In the latter case there is an additional linear term
in the specific heat vs.\ temperature dependence. We suggest that frozen magnetic disorder and
corresponding entropy is responsible for the large overheating at the avalanche-like transition to
the FM state upon increasing the magnetic field.

% >>>>>>>>>>>  SETUP etc.  <<<<<<<<<<<<<<<<<<

\begin{figure}[tb]
\includegraphics{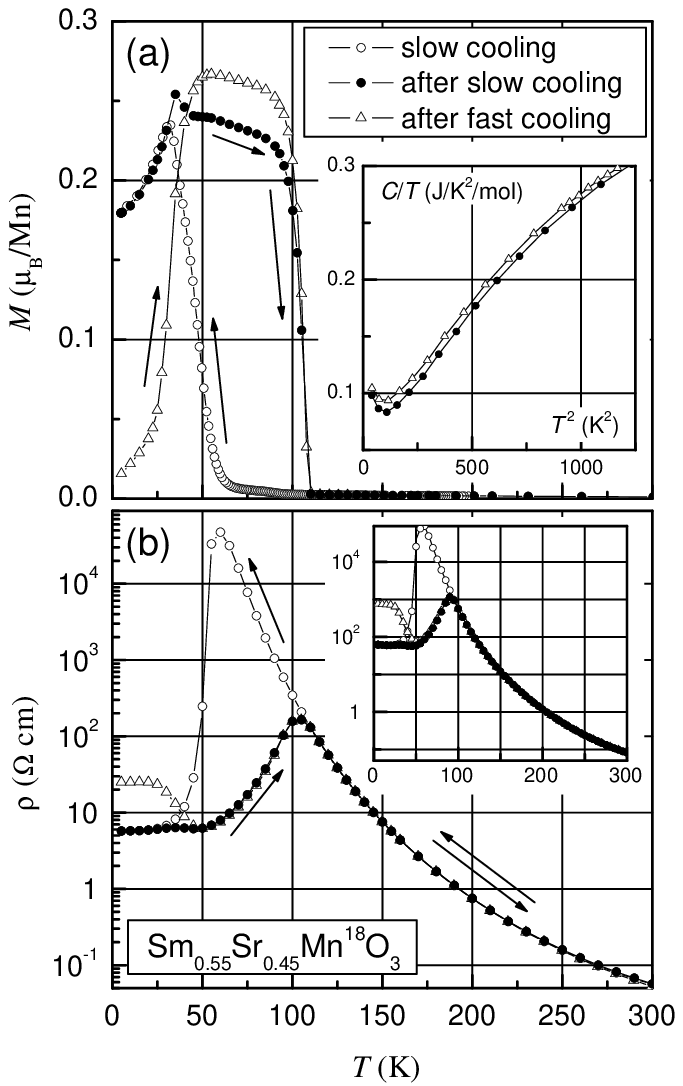}
\caption{\label{f3} The temperature dependences of the low-field (0.01~T) magnetization (a) and
zero-field resistivity (b) of \smsrA\ obtained for different temperature histories: ({\Large
$\circ$}) --- slow cooling, ({\Large $\bullet$}) --- heating after slow cooling, and
($\vartriangle$) --- heating after fast cooling. Insets show the temperature dependence of the
specific heat (a) and resistivity (b) of \ndeusr\ for the same temperature histories.}
\end{figure}

Ceramic \smsrAC\ samples with $x=0.45$, 0.5 and \ndeusr\ were prepared by a solid-state reaction
technique, described in Ref.~\onlinecite{bab17}. The Nd/Eu ratio was selected to fit the Sm ionic
radius. The magnetic and electric behavior of SmSr and (NdEu)Sr samples is qualitatively identical
(see \fref{f3}(b)). The enrichment of the samples by \oeight\ was performed at $T=950^\circ$C and
at a pressure $p=1$~bar for 200~h using the method reported in Ref.~\onlinecite{bab18}.
Magnetization was measured by a Quantum Design MPMS-7 SQUID magnetometer and by a vibrating sample
magnetometer. High-speed (up to 100\,000 samplings per second) measurements of the magnetization
were performed using Fitz's technique and a fast analog-to-digital data acquisition board (Data
Translation). Resistivity, specific heat, and magnetostriction were measured in PPMS-9 cryostat. To
measure magnetostriction we used WK-06-062AP-350 strain gauges (Vishay Intertechnology) bonded to
the sample with a proper epoxy. The striction was detected by the change in resistance of the
strain gauge.

% >>>>>>>>>>>  RESULTS  <<<<<<<<<<<<<<<<<<

% fig.1
\begin{figure}[tb]
\includegraphics{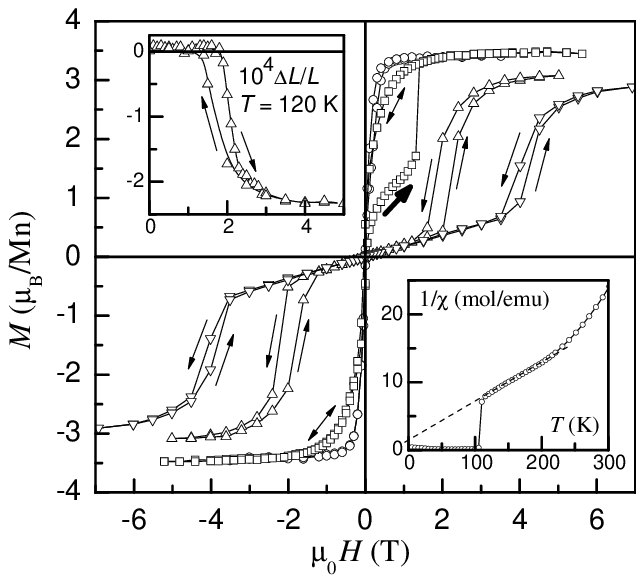}
\caption{\label{f1} Magnetization loops for \smsrA\ recorded after zero-field cooling from room
temperature. The data for the first increase of the magnetic field are shown in the first quadrant
and marked with bold arrows if different from stable curves (III quadrant). $T=5$~K ($\square$),
55~K ({\Large $\circ$}), 120~K ($\vartriangle$), and 140~K ($\triangledown$). The upper left inset
shows magnetostriction at 120~K, the lower right one presents a high-temperature tail of inverse
susceptibility at $\mu_0 H = 0.1$~T.}
% ($\blacksquare$)
\end{figure}

\smsr\ with $x\approx 0.5$ is known to be in the vicinity of the I-M and AFM-FM transition
\cite{MartinPhaseDiags}. So, for this system the electronic and magnetic state can be tuned by the
application of a magnetic field or by oxygen-isotope substitution. \smsrC\ is an insulator in the
low-temperature ground state, and undergoes an I-M transition after \oeight-to-\osix\ substitution,
after reduction of the doping level to $x=0.45$, or under application of the magnetic field of the
order of 1~T \cite{BabSmSr1618}. Our magnetization data (\fref{f1}) show that there are two types
of field-induced transition for \oeight-samples: at low temperatures there appears an irreversible
transition from an AFM to FM state, and at high temperatures $T>T_c$ --- a reversible albeit
slightly hysteretic PM-FM transition. Magnetostriction of about $2\times 10^{-4}$ is clearly seen
to accompany the PM-FM transition (upper inset) in accordance with the data of
Ref.~\onlinecite{TomiokaBandwidth}. The $1/\chi$ data just above $T_c$ show a tendency for AFM
interactions that competes with FM ordering at intermediate temperatures.

% fig.2

\begin{figure}[tb]
\includegraphics{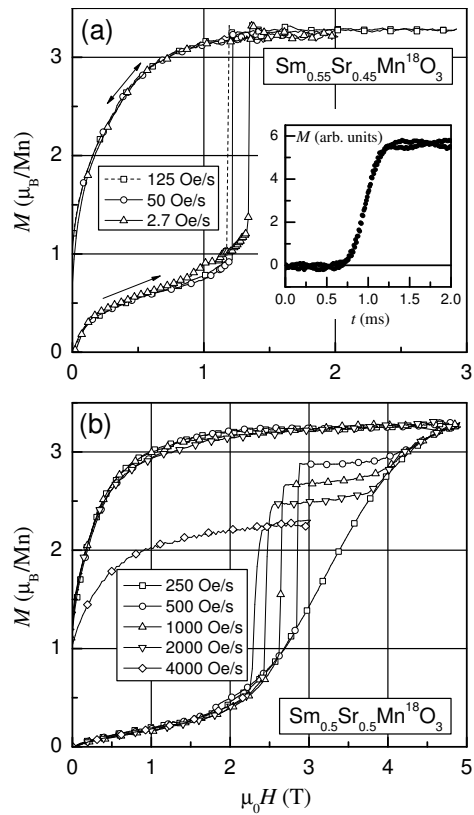}
\caption{\label{f2} The dependence of the magnetization jump on the magnetic field sweep rate for
\smsrA\ (a) and \smsrC\ (b) at $T=5$~K. Inset shows the time dependence of the magnetization during
the jump. This dependence is the same for the sweep rates in the range 200--3000~Oe/s.}
\end{figure}

The irreversible metamagnetic AFM-FM transition is shown to be step-like (\fref{f2}) after
zero-field-cooling. We have observed that the step location depends on the sweep rate of the
magnetic field: the smaller the sweep rate, the larger field value is needed to realize the
transition. For \smsrC\ sample at the rate $\le 250$~Oe/s the transition becomes smooth. This shows
that the step-like transition is not an intrinsic property of a compound. Two question arise at
this point: (i) how sharp are these steps and (ii) is $M$ a function of the magnetic field along
all the `step' or only the start of the transition is triggered by the magnetic field.

To resolve these problems we have studied the $M(H)$ steps at different sweep rates with our
high-speed experimental setup with 10$~\mu$s resolution. The results shown in the inset to
\fref{f2} demonstrate the finite width of the transition of the order of 1~ms. Note that this curve
is the same for different sweep rates (200--3000~Oe/s). Being triggered, the transition will
complete in a definite time independent on the further changes of the $H$. In the case of \smsrC,
this time-scale is approximately 10 times higher (not shown). Thus, our samples with a relatively
small difference in Sr content exhibit a factor of ten difference in the transition time. This fact
cannot be easily reconciled with the scenario of a martensitic transition because the
microstructure of both samples is identical. Moreover, a distribution of avalanches in martensitic
transformations has usually no characteristic time-scales \cite{VivesDistrAval} unlike our
observations.

To further elucidate the origin of such sharp transitions, we have checked the effects of the
cooling rate on the low-temperature magnetic state (\fref{f3}). The quenched-in disorder should be
strongly affected by the cooling rate. In the case of martensitic transformation
\cite{VivesDistrAval}, a slow cooling rate leads to a low defect concentration so that the strains
are released via large avalanches. On the other hand, fast cooling results in a large number of
defects and thus in a sequence of small avalanches. However, as we will discuss below, the behavior
observed in our experiments with different cooling rates shows the opposite tendency. This is in
our opinion a strong argument against the interpretation of the jumps as a consequence of
martensitic phenomena.

Two cooling rates were used in our experiments. In the first, slow cooling, the sample was cooled
at the rate of 1~K/min from 300~K to 5~K in zero applied field for transport and specific heat
measurements and 100~Oe for magnetization experiments. In the second, fast cooling regime, the
cooling rate was 10~K/min for resistivity and specific heat measurements and approximately 20~K/min
for magnetization. Usually, no specific information is provided in the literature on the employed
cooling rate in manganite research, and we assume that most results are obtained using a relatively
fast cooling rate.

% fig.3

Two key points should be noted in the data of \fref{f3}. The first is a huge thermal hysteresis for
the data obtained on heating and cooling, which is generally considered to be a fingerprint of a
first order phase transition. The second is a striking difference in the low-temperature states of
the fast and slow-cooled samples. For the former, we observe a low-magnetization state with a
relatively high resistivity. For the latter, the magnetization is 10 times higher and the
resistivity is smaller. The additional linear term in the specific heat appears for the fast cooled
sample as shown in the inset to \fref{f3}(a). These differences diminish upon heating and
disappears at $T \approx 45$~K.

% fig.4

\begin{figure}[tb]
\includegraphics{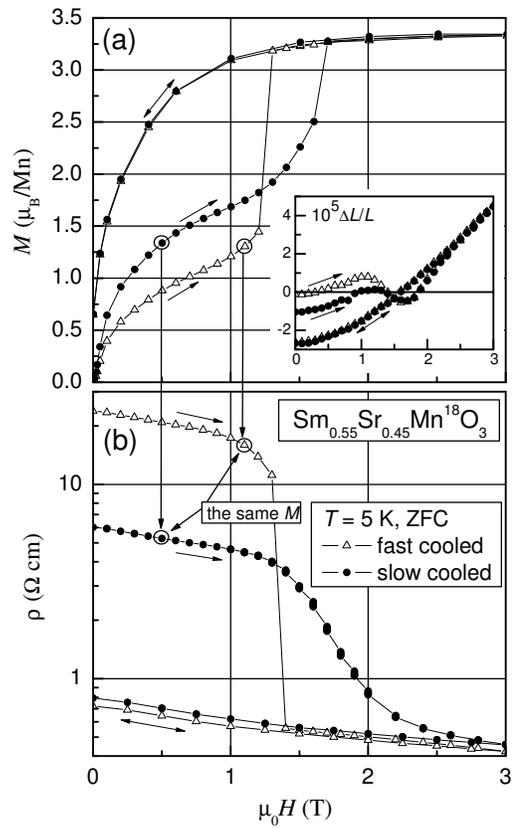}
\caption{\label{f4} Magnetization (a) and resistivity (b) loops for \smsrA\ recorded at 5~K after
slow and fast cooling in zero field, which demonstrate the absence of jumps for slowly cooled
samples. The inset shows a minor effect of the cooling rate on the magnetostriction. Here the
parabolic increase is caused by the magnetoresistance of the gauge itself.}
\end{figure}

The low-temperature magnetization and resistivity loops after slow and fast cooling (\fref{f4})
have qualitatively different behavior. The step-like transition in $M(H)$ and $\rho(H)$ exists only
for fast cooled samples, whereas this transformation is smooth in slow cooled samples. Comparison
of the data of figures~\ref{f4}(a) and (b) shows that the fast-cooled samples have significantly
larger resistivity even at the same value of $M$ (compare, for instance, $\rho(H=0.5, 1.1$~T). This
is evidence for a quenched disorder in fast-cooled samples (cf.
Ref.~\onlinecite{KhomskiiFineMist}). The other evidence of an additional disorder is shown in the
inset to \fref{f3}(a). The fitting of the temperature dependence of the specific heat is the same
for both cooling rates except an additional linear term of the order of 20\% in the fast-cooled
sample. (Note, that conductivity in this case is even lower and `electron' term could not be the
origin of this change). Corresponding extra entropy $\Delta S = 0.25$~J/K/mol is of the same order
as the entropy change upon PM-FM transition ($\sim$0.6~J/K/mol) thus we believe this extra linear
term to have essentially magnetic origin, possibly spin-glass-like type \cite{SGreview}. However,
there is almost no additional strain  (less than $10^{-5}$) in the fast-cooled sample and the
magnetostriction is smooth for the both histories. This, together with the opposite dependence of
the behavior of our samples on cooling rate, mentioned above, in our opinion rules out the
description of these phenomena as a consequence of a martensitic nature of the transition.

Upon cooling, the AFM-FM competition may result in a strongly disordered magnetic state, which has
an excess specific heat and enhanced magnetocaloric effect \cite{ZhitomFrustr}. Ferromagnetic
ordering with the external magnetic field would lead to the reduction of this extra entropy. In
this case, the local release of the frozen entropy may result in the avalanche-like overheating of
the sample because the higher the temperature the more tendency to ferromagnetism is observed at
low temperatures (cf. the data for 5 and 55~K in \fref{f1}). So, both the magnetization and
resistivity change in a jump-like fashion. Such effect was observed recently
\cite{LevyMagnetocalor} on the step-like transition. This scenario assumes the time-scale of the
transition to be inversely proportional to the thermal conductivity of the sample. This is exactly
the case if one compares \smsrAC\ with $x=0.45$~and 0.5. The resistivity of \smsrC\ is at least
four orders of magnitude higher \cite{BabSmSr1618} so that the thermal conductivity should be
lower. We recall that the observed time-scale of the jump for \smsrC\ is 10~times larger than for
\smsrA.

% >>>>>>>>>>>  CONCLUSION  <<<<<<<<<<<<<<<<<<

Summarizing, we studied the magnetization  and resistivity jumps in \smsrAC, especially their
dependence on the cooling rate. The slow (1~K/min) cooled samples demonstrate a smooth AFM-FM
transition upon increase of the magnetic field, in contrast to  the fast ($\sim 10$~K/min) cooled
samples where this transition is step-like with a characteristic time-scale of 1~ms. The results
obtained (the dependence of the jumps on the cooling rate; the constant time-scale of the jumps,
independent on the field-sweep rate; the absence of noticeable magnetostriction) disagree with the
often used interpretation that the magnetization steps originate from the strain release at a
martensitic transition. We suggest that the frozen disorder and the associated entropy is the
origin of an excess specific heat which is released in an avalanche-like way upon applying a
magnetic field, resulting in a `heat burst' in the sample and in steps in the magnetization and
resistivity.

This study was supported by the Netherlands Organization for the Advancement of Pure Research (NWO)
and INTAS project (01-2008).

\end{document}